\documentclass[pra,twocolumn]{{revtex4}}
\usepackage{amsmath,amssymb,graphicx,hyperref}

\newcommand {\ket}[1] {|#1 \rangle}
\newcommand {\braket}[2] {\langle #2 | #1 \rangle }
\newcommand {\densop}[2] {|#1 \rangle \langle #2 |}

\newcommand {\tit}[1] {\textit{#1}}
\newcommand {\tbf}[1] {\textbf{#1}}
\newcommand {\trm}[1] {\textrm{#1}}

\newcommand{\eqn}[1]{\begin{eqnarray} #1 \end{eqnarray}}

\begin{document}

\title{A relational approach to quantum reference frames for spins}
\author{Jacques Pienaar}

\begin{abstract}

In the literature on quantum reference frames, the internal (relative) properties of a system are defined as those which are preserved under an arbitrary change of reference frame. For a system of quantum spins, these are all properties preserved by proper spatial rotations of the laboratory. However, this approach does not account for the hypothetical possibility of the laboratory becoming entangled to the system, as described by a second laboratory (the `Wigner's friend' scenario), in which case the relationship between the two laboratories is not a rotation, but is fundamentally quantum. To overcome this limitation, we re-define the reference frame transformations to be those that preserve the fidelities between subsystems. This enables us to derive U(2) as the correct symmetry group for transformations of a system of $N$ spin-half particles. Next, we propose that systems having the same internal properties should be regarded as physically equivalent in the absence of an external frame. Remarkably, this implies that a single spin in a superposition relative to a spin magnet is equivalent to a macroscopic superposition of the magnet relative to the spin. We discuss the implications of this result for the Wigner's friend paradox.

\end{abstract}

\maketitle

\section{Introduction}

What is the meaning of a quantum state $\ket{\Psi}$? It can ultimately only have meaning in the context of a laboratory in which it is possible to prepare and measure quantum systems under controlled conditions. The state represents a comparison between the state of the observed system and the states of the measuring devices. If we assume the physical state of the laboratory is fixed, then any change in the state of the system must be interpreted as a physical change in the system (and not the laboratory). More generally, the laboratory can itself be treated as a physical system subject to change, and hence changes in the state of the system only represent changes in the relationship between system and laboratory. This is called a \tit{relational} interpretation of the state. 

Relational interpretations are not new; they include Everett's `relative state interpretation' \cite{EVE}, Rovelli's relational interpretation \cite{ROV}, and Quantum Bayesianism \cite{QBISM}. These approaches all agree that the quantum state describes relative information, but disagree on \tit{whose} information it is, and \tit{what} the information is about. In all cases, however, the measurable properties of a system can be split into two categories: those that change when the system is measured in a different laboratory, and those that are measured to be the same in all laboratories. We call these properties \tit{external} or \tit{internal}, respectively \footnote{The terminology `external' and `internal' has unfortunately been used quite differently in other contexts. For example, it is common to refer to the spin of a single electron as an internal degree of freedom, although by the present definition it is external, since it must be referred to a reference spin, like a large magnet. The reader should therefore take care not to conflate the terms as used here with their usage in other contexts).}. 

At one extreme, a \tit{subjective Bayesian} (see eg. Ref. \cite{FUCHS2}) might say that the quantum state represents only the beliefs of a rational agent. In this case all properties are external, since every state is a valid description of the beliefs of some agent, and the laws of physics describe how these beliefs should be updated in light of measurement outcomes. At the other extreme, a \tit{psi-ontologist} (see eg. Ref. \cite{LEI}) would claim that two different states represent two different objective realities, hence all properties of the system are internal and do not depend on the observer.

In this work, we wish to explore a middle ground, inspired by Rovelli's relational interpretation, according to which different observers may give different accounts of the same physical situation. It follows that any disagreements between them must be about external quantities. However, since the observers themselves must obey quantum mechanics, their accounts must also be consistent with each other to some degree. The internal properties are precisely those facts that all observers can agree upon. Our aim is to pin down what these properties are, and to decide exactly what it means for two states to represent the `same physical situation' in a relational quantum world.

It is clear that the distinction between external and internal properties cannot be fundamental, so long as we give no fundamental significance to the attendant distinction between the laboratory and the system. Nevertheless, the distinction is relevant to practical considerations. For example, consider two laboratories attempting to exchange quantum information, but whose measuring devices are not mutually aligned. Since the receiver is ignorant of the sender's reference frame, he must sum over all possibilities, in effect applying a random transformation to the external degrees of freedom. This destroys any information encoded in these degrees of freedom, reducing the parties' ability to communicate. This is the basis for the literature on \tit{quantum reference frames} \cite{POU,BART,BART2}, which provides an extensive toolbox of methods for performing quantum communication, cryptography, teleportation and other information-theoretic tasks in situations where the laboratories do not share a given reference frame.
 
The work on quantum reference frames proceeds from the assumption that the separation between internal and external degrees of freedom is known. Usually, it is given in the form of a symmetry group $\mathcal{G}$ that is postulated to preserve only the internal degrees of freedom. The symmetry group is justified by appealing to some natural symmetry of the physical setting in question. For example, if the systems to be communicated are spin-half particles, it seems logical that the measuring instruments in the laboratories are Stern-Gerlach magnets or similar. These instruments indicate a preferred \tit{Cartesian reference frame} for the laboratory, representing the labelling of the $(x,y,z)$ co-ordinates in the lab. Since the Cartesian reference frames of the laboratories are related by a proper rotation, the associated symmetry group is SO(3). For quantum particles, we typically seek a unitary representation of the rotation group, hence $\mathcal{G} = \trm{SU(2)}$. 

Most of the literature to date has restricted attention to transformations that preserve the separation between system and laboratory. That is to say, if one laboratory performs a measurement on the system, the state is updated according to the outcome of this measurement \tit{for all laboratories}, not just the laboratory in question. This assumption excludes the possibility of treating the laboratories themselves as quantum systems relative to one another. For example, consider the following thought experiment due to Wigner \cite{WIG}: an experimenter called Fran (a friend of Wigner) performs a measurement of the system and obtains an outcome. It should then be possible for another experimenter in a different laboratory (Wigner) to treat Fran and her laboratory as a quantum system, and the measurement as a physical interaction that entangles the state of Fran to the state of the system. This physical interaction is a transformation that applies not just to the system alone, but also to the whole laboratory of Fran, thereby including this laboratory and Fran herself as part of the `system'. Thus, Fran's laboratory is in a superposition relative to Wigner's laboratory. So far, there have been few attempts to apply the tools of quantum reference frames to physical settings of this more general type. 

We note the following exceptions. Palmer et. al. \cite{PAL14} have considered the task of changing reference frames in a setting where the reference frame is an explicitly quantum object; however they do not extend their analysis to `Wigner's friend'-type scenarios. Angelo et. al. \cite{ANG11,ANG12}, and Pereira and Angelo \cite{PER15}, following work by Aharonov et. al. \cite{AHA1}, consider the problem of treating quantum systems as observers and changing frames between them. However, that work is concerned with linear position and momentum degrees of freedom, whereas we consider angular momentum and spin (since the rotation group is compact, we can apply the standard tools of quantum reference frames). Finally, H\"{o}hn et. al. \cite{HOH15a, HOH15b, HOH15c} have managed to reconstruct quantum mechanics and its key symmetry groups for qubits using an operational approach that inspired the approach used in the present work; however they also do not address the Wigner's friend scenario.

In this work, we make an incremental step towards extending the formalism of quantum reference frames to more general scenarios of the kind applicable to the Wigner's friend thought experiment. Our main idea is to try and derive the symmetry group of reference frame transformations from first principles, rather than postulating it from a classical limit, in the hopes that the resulting physical symmetry will permit us to draw an equivalence between the apparently irreconcilable states seen by Wigner and his friend. Although we ultimately do not succeed in this task, we are able to find an example that transcends the usual system-apparatus division in the absence of an external reference frame, indicating that future efforts along similar lines might bear fruit.

The outline of the paper is as follows. In Section \ref{Sec:Basic} we consider a system composed of spins and ask what properties should be the same for all observers. We argue that all observers should be subject to the same constraints on their ability to distinguish between subsystems of the total system. Taking the \tit{quantum fidelity} between subsystems as a measure of their distinguishability, we obtain U(2) as the symmetry group that preserves them. In Section \ref{Sec:Finale} we examine the bearing this result has on the microscopic / macroscopic distinction and the Wigner's friend thought experiment. Section \ref{Sec:End} contains our conclusions and outlook.

\section{The internal properties of spins \label{Sec:Basic}}

Our task is to define the internal properties of general pure states $\ket{\Psi}$ of $N$ spins, namely, the attributes of the system that do not depend on the observer's laboratory. In what follows, we assume that all laboratories can agree on a labeling of the $N$ spins, so that if one observer specifies a subsystem of spins, say ``spins numbered $7-12$", all the observers agree on which subsystem this refers to. Measurements of spin performed within a laboratory are relative to the (arbitrary) orientation of that laboratory's spatial axes. The procedure for measuring `spin up' in a given direction is assumed to be the same for all spins, hence labeling the `up' direction for one spin fixes the label for all spins in the same lab. Finally, we assume all observers can measure the number of spins in any subsystem without difficulty. Other than that, the devices in the laboratories are not calibrated with each other. 

The problem can now be understood as follows. If a system of $N$ spins is prepared according to a fixed preparation procedure, it will nevertheless have a different state according to different laboratories. We want to know: how much can the states seen by two laboratories differ? Clearly, any differences can only be attributed to a change of laboratory frame, since the preparation of the system is invariant. We have assumed that the laboratories are related by collective transformations of their apparatuses that preserve the number of spins and the spin ordering, but otherwise the symmetry group that relates the laboratories is unknown.  

Our first clue comes from asking how well the laboratories are able to distinguish different preparation procedures. Imagine that a machine prepares $N$-spin systems according one of two preparation procedures, $A$ or $B$. In the first phase of the experiment, it prepares many systems using procedure $A$ and sends multiple copies to each laboratory for analysis. In the second phase, it does the same but only using procedure $B$. As a result, each laboratory is able to perform tomography on the systems they received and determine the state obtained from each preparation procedure relative to their own instruments. In general, the pair of states $\rho_A,\rho_B$ seen by one lab will not be the same as the pair $\rho'_A, \rho'_B$ in another lab. 

In the third phase of the experiment, the machine sends only a single copy of the system to each laboratory, and in each case it chooses randomly which preparation procedure to use: with probability $p$ it applies procedure $A$, and with probability $(1-p)$ it applies $B$. Each lab is then asked to guess whether the system they received was prepared using procedure $A$ or $B$. They can use only the knowledge they have gained in the previous two phases, plus the value of $p$.    

Since the pair of states being distinguished will in general differ among laboratories, it is \tit{a priori} possible that one laboratory might consistently perform better at the task than another laboratory (for instance, if the pair of states seen by one laboratory are more alike than the states as seen by another laboratory). Hopefully, this possibility strikes the reader as rather unnatural. After all, the laboratories are all trying to distinguish systems produced by the \tit{same} pair of preparation procedures, and these procedures are frame-independent notions, so we intuitively expect that no laboratory has any special advantage over another. Hence, the likelihood of succeeding should be \tit{independent} of which laboratory (i.e. observer) is asked to perform the task. 

Without loss of generality, we can imagine that the machine implements the two preparations by first preparing a single system according to a fixed procedure and then choosing one of two $N$-spin subsystems, designated $A$ or $B$, as the output. For example, let $N=3$. Starting with an initially prepared state of $M \geq N$ spins, the machine could choose either spins $5,11,14$ (subsystem $A$) or spins $17,42,99$ (subsystem $B$), discard the rest, and re-label the chosen spins $1,2,3$. In this way, the choice of preparation procedure reduces to a choice of subsystems sampled from a larger system with a fixed preparation. With this perspective, the considerations of the previous paragraph amount to the statement that all laboratories should be equally good at distinguishing a given pair of subsystems of a total system prepared by a fixed procedure. This forms our first postulate:\\

\tbf{Postulate 1:} Let $A$ and $B$ label two (not necessarily disjoint) equal-sized subsystems of a quantum system prepared by a fixed procedure. We assume that all observers have access to identically prepared copies of this total system and its subsystems. Then the ability of any observer to distinguish $A$ from $B$ (when presented with a random sample of either one) must be the same for all observers. \\

(Note that although we motivated this postulate from the example of spins, its scope is quite general). For a given pair of states $\rho_A$ and $\rho_B$, quantum mechanics supplies bounds on their distinguishability. If just a single copy of the state is provided, then the appropriate measure of distinguishability is the \tit{quantum error probability}. More generally, after randomly choosing whether to prepare $A$ or $B$, the machine could send a fixed number of copies of the system to each lab. In this case, there is a whole zoo of distinguishability measures that could be used (see eg. Ref \cite{FUCHS}). We will adopt the \tit{quantum fidelity} as the measure of interest in the remainder of this work, but we conjecture that the results obtained here hold also for other measures of distinguishability. The quantum fidelity can be expressed as: 
\eqn{ \label{eqn:fidel}
F(\rho_A,\rho_B) :=  \underset{ \{ F_k \} }{\trm{min}}  \, \sum_k \,\sqrt{ \trm{Tr} \left[ \rho_A F_k \right] \trm{Tr} \left[ \rho_B F_k \right]  } \, ,
}
where the minimisation is over all possible POVMs $\{ F_k \}$ on the Hilbert space of the systems. 

Why choose the fidelity? For spins, the fidelity can be related to the relative spatial angles between the spin vectors. Specifically, if we restrict attention to a tensor product of $N$ pure spin states, then the state space is isomorphic to a set of unit vectors in 3-dimensional space (Bloch vectors). In that case, a natural candidate for the internal properties of the system are the relative angles between the Bloch vectors. The angle $\theta_{ij}$ between the $i_{\trm{th}}$ and $j_{\trm{th}}$ vectors is directly related to the overlap $|\braket{\phi_i}{\phi_j}|$ between their respective states, which is just the fidelity in this case. For more general (entangled) states, the fidelities can be related to angles between Bloch vectors of the purifications of the spin states (Uhlmann's theorem, Ref. \cite{UHL}). The connection between the fidelities and the relative angles between spatial vectors is one of the main inspirations for using the fidelity as the measure of distinguishability. However, it is merely a heuristic motivation, as the Bloch vectors of higher-dimensional systems obviously do not exist in 3-dimensional space. Ultimately, we have chosen the fidelities mostly because they are mathematically convenient and lead to interesting results.

An $N$ spin system can be partitioned into subsystems of different numbers of spins. Obviously, if the subsystems have different sizes, they are perfectly distinguishable by measuring the number of spins, so the only nontrivial cases arise when the subsystems have equal numbers of spins. This is important, because the fidelity is defined only for subsystems of equal size. Using the fidelity, we can now give a mathematical version of Postulate 1 for spins:\\

\tbf{Postulate 1 for spins:} Given an arbitrary pure state $\ket{\Psi}$ consisting of $N$ spins, the \tit{internal properties} are the set of fidelities $F(\rho_A,\rho_B)$ between all subsystems $\rho_A,\rho_B$ with equal numbers of spins. \\

We next show that these fidelities are preserved by (and only by) actions of the unitary group U(2):\\

\tbf{Theorem 1:} The set of fidelity-preserving transformations acting on the $N$-spin Hilbert space is the image of a unitary representation of the group U(2), namely:
\eqn{
\pi: V \in \trm{U(2)} \mapsto V^{\otimes N} \in \trm{U($2^N$)} \, .
}

\tit{Proof:} The action of a unitary $U$ on $\ket{\Psi}$ induces the following transformations on the reduced states:
\eqn{
\rho_A \rightarrow \trm{Tr}_{\neq A} \left[ U \densop{\Psi}{\Psi} U^{\dagger} \right] :=  \mathcal{E}(\rho_A) \nonumber \, , \\
\rho_B \rightarrow \trm{Tr}_{\neq B} \left[ U \densop{\Psi}{\Psi} U^{\dagger} \right] :=  \mathcal{E}'(\rho_B) \, .
}
It can be shown using Uhlmann's theorem \cite{UHL} that the fidelity will be preserved only if the CPT maps are the same, i.e. $\mathcal{E}'=\mathcal{E}$. We then make use of a theorem due to Molnar \cite{MOL}, which states that the fidelities are preserved if and only if $\mathcal{E}(\rho) =  V \rho V^{\dagger} $ for some unitary $V$ acting on the Hilbert space of the subsystems (this is a generalisation of a well-known theorem of Wigner \cite{WIG2}). It follows that $U$ decomposes into $U= U_A U_B U_R$ where $R$ labels the systems not in $A$ or $B$, and $U_A = U_B = V$. Applying this to all pairs of single-spin subsystems leads to the conclusion that $U$ preserves the fidelities iff $U=V^{\otimes N}$ with $V \in \trm{U(2)}$. Under the map $V \mapsto V^{\otimes N}$, the image of U(2) in U($2^N$) is therefore precisely the set of fidelity preserving maps acting on the space of $N$ spins. $\Box$ \\

While U(2) is the appropriate group for states in Hilbert space, one is typically concerned only with `physical states' in the form of projectors $\densop{\Psi}{\Psi}$, which live in the projective Hilbert space. In that case, the following corollary applies: \\

\tit{Corollary:} Under the representation mentioned in Theorem 1, fidelity-preserving maps that act on the projective Hilbert space constitute the image of the group SO(3). \\

\tit{Proof:} Theorem 1 implies that the fidelity-preserving unitaries acting on the projective space have the form $W^{\otimes N}$, where $W$ is any element of the 1-spin projective unitary group PU(2). This defines the representation of PU(2) by fidelity-preserving unitaries acting on the projective space, and since PU(2) is isomorphic to SO(3), the statement of the Corollary follows. $\Box$ \\

The symmetry group of reference frame transformations is usually taken for granted in the literature, but we have derived it using a novel assumption (Postulate 1). Note that since the fidelities are not in general independent, they should be reducible to a smaller set of independent internal properties, which we have not attempted to identify here. It is straightforward to do so using tools from the literature on quantum reference frames; for example, Eq. 3.20 in \cite{BART} presents the decomposition into collective and relative (i.e. internal) degrees of freedom for the group SU(2), and one sees that the internal degrees of freedom span the degenerate subspaces of states with definite total angular momentum. 

Although Theorem 1 states that fidelities are guaranteed to be preserved by all and only those unitaries with the form $U=V^{\otimes N}$, this does not imply that two states having the same fidelities must necessarily be related by a unitary map of this form. An example of this intriguing situation will be explored in the next section. 

\section{Physical interpretation \label{Sec:Finale} }

In analogy with special relativity, let us assume that transformations that preserve the internal properties relate physically equivalent situations. That is, the internal properties constitute the \tit{observer-independent} core of quantum mechanics: \\

\tbf{Postulate 2:} In the absence of a preferred external reference frame, two quantum states of the same number of spins represent the \tit{same physical situation} iff they have the same internal properties (i.e. the same fidelities between their subsystems). \\

This leads to the following counter-intuitive result: a microscopic superposition as seen by one lab can appear to be a macroscopic superposition as seen by a different lab. Let $\mathcal{B} := \{ \ket{\eta_1, \eta_2, ... \eta_M} \}$ with $\eta \in \{0,1 \}$ be the computational basis for a set of $M$ spins. Consider the following two states, expressed in the basis $\mathcal{B}$:
\eqn{
 \ket{\phi} &=& \ket{00...0} \left( \alpha \ket{0}+ \beta \ket{1} \right) \nonumber \\
&:=& \ket{0}_1 \ket{\Phi} \, , \nonumber \\
 \ket{\phi'} &=& \left( \alpha \ket{00...0}+ \beta \ket{11...1} \right) \ket{0} \nonumber \\
&:=& \ket{\Phi'}  \ket{0}_M \, .
}
The first state $\ket{\phi}$ might represent, for example, a large magnet consisting of $M-1$ spins, with the $M_{\trm{th}}$ spin in a superposition relative to the magnet. The basis $\mathcal{B}$ is a natural or \tit{preferred} choice of labeling in a laboratory where the physical states corresponding to elements of $\mathcal{B}$ are easy to prepare, and where superpositions of these elements are easy to prepare whenever the corresponding elements of $\mathcal{B}$ are similar to each other. An example is the state $\ket{\phi}$, whose superposed states differ at only a single spin in the basis $\mathcal{B}$; we call $\ket{\phi}$ a \tit{microscopic} superposition relative to such a lab. Conversely, if the two superposed states are very different when expressed in the natural basis of the lab, the state represents a \tit{macroscopic} superposition. Thus $\ket{\phi'}$ is macroscopic relative to the lab for which $\mathcal{B}$ is the natural basis. If we have a preferred basis $\mathcal{B}$ in which to prepare both these states, we will find they appear physically different, the state $\ket{\phi'}$ being significantly harder to prepare than the state $\ket{\phi}$. We now observe that these two states have the same internal quantities, namely:
\eqn{
F(\rho_A,\rho_B) &=& 1, \, \, \, \, \trm{if $A,B$ exclude spin $M$;} \nonumber \\
F(\rho_A,\rho_B) &=& |\alpha|^2 \, \, \, \, \trm{if $M$ appears in one of $A,B$;} \nonumber \\
F(\rho_A,\rho_B) &=& |\alpha|^4 \, \, \, \, \trm{if $M$ appears at different} \nonumber \\
&& \qquad \quad \, \trm{sites in both $A,B$;} \nonumber \\
F(\rho_A,\rho_B) &=& 1 \, \, \, \, \trm{if $M$ appears at the same} \nonumber \\
&& \qquad \quad \, \trm{site in both $A,B$} \, .
}
Postulate 2 then asserts that, \tit{in the absence of a preferred external reference frame}, these states should be physically equivalent. To see how this can occur, consider a different basis $\mathcal{B}'$ related to $\mathcal{B}$ by a transformation that exchanges $0 \leftrightarrow 1$ in all elements except for $\ket{00...0}$ and $\ket{11...1}$. For instance, for $M=3$ the superposition $(\ket{000}+\ket{001})/\sqrt{2}$ in the basis $\mathcal{B}$ corresponds to $(\ket{000}+\ket{110})/\sqrt{2}$ in the basis $\mathcal{B'}$. Notice that $\ket{\phi}$ and $\ket{\phi'}$ switch roles when expressed in terms of basis $\mathcal{B}'$. This means that if they are both prepared in a lab for which $\mathcal{B}'$ is the preferred basis, $\ket{\phi'}$ will be microscopic (hence easier to prepare) and $\ket{\phi}$ will appear macroscopic. This example clearly shows that if no preferred basis is specified (i.e. there is no external lab to tell us which basis is most natural) the states $\ket{\phi'}$ and $\ket{\phi}$ may be regarded as representing the same physical situation. 

\tit{Comment:} there is a certain intuition according to which one might expect an electron in superposition with respect to a large magnet to `see' the magnet as being in a superposition. This procedure of `stepping into the electron's shoes' and talking about `what the electron sees' seems to fit with the transformation from $\ket{\phi}$ to $\ket{\phi'}$. One must be careful in making such claims, however, since as we have shown above, this physical equivalence of frames can only be maintained in the absence of a preferred external basis. In practice, there always exists a preferred basis for the external laboratory, which is singled out by decoherence of the system and measuring apparatus by a common environment. This is why, for example, placing a single particle in a spatial superposition over a long distance is considered `macroscopic': the environment tends to decohere systems in the position basis. An interesting task for future work would be to include the environment in the preceding analysis, to see explicitly how it breaks the symmetry between the states $\ket{\phi'}$ and $\ket{\phi}$.

Setting aside decoherence, one might still expect this surprising equivalence of states to have some bearing on the problem of Wigner's friend. After all, it presents a dual perspective that seems to correspond to the conflicting experiences of the two parties: in one point of view (Wigner), the $M_{\trm{th}}$ spin is in a superposition, while in the other point of view (Fran) it's spin has a definite value. The fact that the two points of view are reconciled by Postulate 2 seems promising. Alas, further investigation leads us to conclude that the analogy cannot be made precise. In order to do so, it would be necessary to treat Fran as a physical system (perhaps herself composed of spin-half particles) initially entangled to spin $M$. The task would then be to discover a state having the same fidelities as the original, but in which the spin $M$ has a definite value, in accordance with what we expect Fran to see. The problem is that the equivalence of $\ket{\phi'}$ and $\ket{\phi}$ seems to be rather a special case: simple generalizations, like adding more spins to the initially superposed state, or even adding an extra magnet pointing in a different direction, lead to states whose fidelity-equivalent counterparts don't seem to possess the desired properties. For the present, it is left as an open problem to classify the sets of physically equivalent states induced by the fidelities via Postulate 2, but our efforts so far seem to indicate that these equivalence classes are not rich enough to capture the duality between Wigner and Fran's conflicting experiences. 

Despite this disappointment, the example does show that if we take internal properties as fundamental, and their symmetry transformations as merely derived, then the line between apparatus and measured system can be transgressed. \\

\section{Conclusions \label{Sec:End} }

We have shown that it is possible to derive U(2) as the group that preserves the internal degrees of freedom of spins by postulating (cf. Postulate 1) that the internal degrees of freedom are the fidelities between subsystems of spins. Furthermore, we proposed that states having the same internal quantities should represent the same `physical situation', differing only in terms of the description relative to some laboratory (Postulate 2). This led to the observation that, in absence of a preferred external laboratory, microscopic and macroscopic superpositions become equivalent.

The work presented here was inspired by the idea that, in certain circumstances, a single quantum particle can be considered an observer. Indeed, the approach shows promise for reconciling the viewpoints of Wigner and his friend in the thought experiment, by utilizing the notion of an equivalence class of states all having the same internal properties. Unfortunately, we were unable to achieve this goal using the fidelities as the relevant degrees of freedom.  

The author thanks \v{C}aslav Brukner, Philipp H\"{o}hn, Alexander Smith and an anonymous referee for helpful comments. This work has been supported by the European Commission Project RAQUEL, the John Templeton Foundation, FQXi, and the Austrian Science Fund (FWF) through CoQuS, SFB FoQuS, and the Individual Project 2462.


\end{document}